%% file: cache.tex
\documentclass[12pt]{article}

\usepackage{sbc-template}

\usepackage{graphicx,url}

\usepackage[english]{babel}   
\usepackage[utf8]{inputenc}

\newcommand{\eat}[1]{}

\usepackage{xspace}
\usepackage{amsmath}
\usepackage{amsthm}

\usepackage{amssymb}
\usepackage{amsfonts}
\usepackage{comment}
\usepackage{bm}
\usepackage{comment}
\usepackage{graphics}
\usepackage{graphicx}
\usepackage{listings}
\usepackage{url}
\usepackage{lscape}
\usepackage{mathrsfs}

\usepackage{graphicx}
\usepackage{latexsym}

\usepackage{xspace}
\usepackage{amsmath}
\usepackage{amssymb}
\usepackage{amsfonts}
\usepackage{comment}
\usepackage{bm}
\usepackage{comment}
\usepackage{graphics}
\usepackage{graphicx}
\usepackage{listings}
\usepackage{url}
\usepackage{lscape}
\usepackage{mathrsfs}

\usepackage{cite}
\usepackage{graphicx}
\usepackage{verbatim}
\usepackage{url}
\usepackage{mathrsfs}
\usepackage{amsmath}

\usepackage{graphicx} 

\newcommand {\omegavec}{{\mbox{\boldmath $\omega$}}}
\newcommand {\upsilonvec}{{\mbox{\boldmath $\upsilon$}}}

\newcommand {\beq}{\begin{equation}}
\newcommand {\eeq}{\end{equation}}
\newcommand {\barr}{\begin{array}}
\newcommand {\earr}{\end{array}}
\newcommand {\bearn}{\begin{eqnarray*}}
\newcommand {\eearn}{\end{eqnarray*}}
\newcommand {\bear}{\begin{eqnarray}}
\newcommand {\eear}{\end{eqnarray}}

\renewcommand{\marginpar}[1]{}

\newtheorem{proposition}{Proposition}[section]

\newtheorem{definition}{Definition}[section]

\newtheorem{example}{Example}[section]
\newlength{\labelexample}
\title{Enabling Information Centric Networks through Opportunistic Search, Routing and Caching \thanks {This research was supported in part by grants from CNPq and FAPERJ.} }

\author{Guilherme Domingues\inst{1}, Edmundo de Souza e Silva\inst{1}, Rosa Leão\inst{1}, Daniel S. Menasché\inst{1} }

\address{Universidade Federal do Rio de Janeiro
  (UFRJ)\\
  Rio de Janeiro, RJ -- Brasil
}

\begin{document} 

\maketitle

\begin{abstract}
Content dissemination networks are pervasive in todays' Internet.  
Examples of content dissemination networks include peer-to-peer networks (P2P), content distribution networks (CDN) 
and information centric networks (ICN). In this paper, we propose a new system design for information centric networks which 
leverages opportunistic searching, routing and caching. Our system design is based on an hierarchical tiered structure.
Random walks are used to find content inside each tier, and gateways across
tiers are used to direct requests towards servers placed in the top tier, which are accessed in case content replicas are not found in lower tiers.
Then, we propose a model to analyze the system in consideration.  The model yields metrics such as mean time to find a content and 
the load experienced by custodians as a function of the network topology.  Using the model, we identify tradeoffs between these two metrics, 
and numerically show how to find the optimal time to live of the random walks.
\end{abstract}

\eat{ \chapter*{Acknowledgements} }

\input{intro1}

\input{literature}

\input{system}

\input{analysis}
\input{experiments}
\input{conclusion}

\bibliographystyle{sbc}


\bibliography{cache}

\end{document}

%% file: intro1.tex
\section{Introduction}

\marginpar{actual scenario} In today's Internet, there is a
strong demand for content dissemination networks, such as \emph{social networks}  
(e.g.  Facebook) and \emph{video networks} (e.g. Youtube and  
\eat{ Netflix ~\cite{netflix}, Hulu ~\cite{hulu}, }RIO~\cite{rio06}). 
\eat{ and \emph{collaboration networks} (e.g the encyclopedia Wikipedia) . } To support this 
growing demand, two broad classes of solutions, supported by IP host-to-host 
communications, have been proposed: Content Delivery Networks (CDNs) and Peer-to-Peer Networks (P2P).

\marginpar{cdn} In CDNs, \emph{dedicated servers} store all the information
published. Copies of popular contents are placed close to the users within cache servers. 
Users are automatically and transparently re-directed to the most appropriate server
by a central authority. Quality of Service (QoS) and advanced monitoring 
techniques are deployed, through proprietary solutions, to redirect each user requests  
(e.g., Akamai Networks). \eat{Mirror Image Networks ~\cite{mirrorimage},  Limelight Networks  ~\cite{limelight}}
In CDNs, a set of \emph{origin servers} store a copy of all published content.
Popular contents are also stored in \emph{cache servers} close to users, so as to minimize the service delay 
experienced by the requesters. Requests for content are sent through IP routers 
to the central authority. Content flows back to users, along the IP routers.

\marginpar{p2p} In P2P systems, \emph{peers} act as both client and servers for contents. Peers' requests are sent to other
peers, from where content fragments  (\emph{chunks}) can be retrieved.  Bittorrent and Emule are examples of P2P systems. 
As soon as peers have downloaded all desired content chunks, they may either
leave the system or remain as future 
providers for chunks (\emph{seeders}).  
 While seeders, P2P nodes can leave the system for different reasons (\emph{e.g.}, closing their connection to other peers, power failures, mobility). Indeed, there are no guarantees on the time 
a seeder will remain present in a P2P system.   If peers leave the system immediately  after concluding their downloads, content might become unavailable which might lead to instability~\cite{selfsust}.   Peers exchange information
through IP routers, but in contrast to CDNs, 
there are no dedicated servers to cache content close to users.

\marginpar{future scenario} According to ~\cite{dirkkutscher}, global traffic in the Internet will increase by a factor of four from 2009 to 2014,
approaching 64 exabytes per month in 2014, compared to approximately 15 exabytes per month in 2009. Global mobile
data traffic is expected to double every year through 2014. Despite the tremendous success of CDNs and P2P systems, they present issues if we consider the
\emph{scalability} and \emph{reliability} envisioned for next generation of
content dissemination networks. With billions of mobile nodes claiming for
contents, central decision policies at CDNs tend not to scale. In P2P networks,
the absence of
dedicated cache servers close to users, as well as the lack of guarantees for a
peer to remain in the system after obtaining the desired content, hampers reliability.

\begin{figure}[h!]
\center
\includegraphics[scale=0.30]{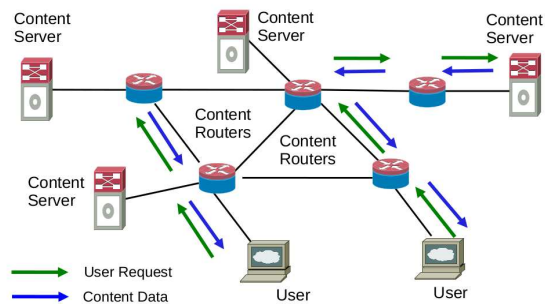}
\caption{ICN scenario}
\label{icnscenario}
\end{figure}

The scalability and reliability challenges for massive distribution content
gave rise to a new research area: Information Centric Networks (ICN) ~\cite{scott}. In ICNs, users know the \emph{name}
of a content being searched, before issuing requests. All searched content is previously stored in the network through subscriptions to the network. 
Caching at all routers, also referred to as \emph{universal caching}, is considered.
 Under universal caching, caching  is provided to all users, and can be potentially implemented by all routers, 
being pervasive along the entire network. 
Figure~\ref{icnscenario} exemplifies this scenario. Questions on how to deploy efficient algorithms to explore \emph{universal caching} 
and on how to define \emph{inter-domain} routing policies give rise to important challenges.

\begin{figure}[h!]
\center
\includegraphics[scale=0.28]{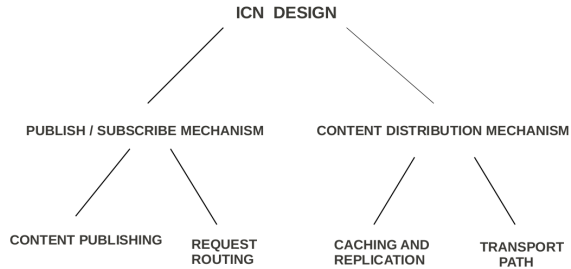}
\caption{Publish/subscribe and content distribution mechanisms}
\label{icntaxonomy}
\end{figure}

Figure~\ref{icntaxonomy}  shows some of the most important ICN features.
In ICNs, content is published without any explicit destination address.
Receivers subscribe to the network through a query (request) for named contents.  Users' 
requests are forwarded by structured or non-structured topologies.   When there is a matching between
a query and a stored content, content is delivered to the subscribers.  During delivery, content might be cached 
and replicated in the network.  
This strategy is known as \emph{in-networking caching}.  
Content can be sent to subscribers over TCP/IP transport mechanisms over paths which are oblivious to the path taken by the requests. 
Alternatively, content can be sent
in the reverse path of requests trails stored across the caches. We reference caches in ICNs henceforth as \emph{cache-routers}.

In this article, our main contribution is a novel ICN architecture with routers disposed along hierarchical tiers domains, without a global
lookup service to map names to IP addresses to all content present in the
network:

$\bullet$ \textbf{System design: } We propose an opportunistic search algorithm 
to be  executed across hierarchical tiers.  Random walks are used to opportunistically find replicas of the content inside a tier.
If no replicas are found, requests are forwarded to another tier in the direction of publish area servers, which maintain fixed copies of the content. 
 This way, we cope with the tradeoff between 
exploration of new routes to content replicas and exploitation of known routes to fixed replicas.   
In addition, we also propose an opportunistic caching policy using reinforced counters. The content placement and eviction
policies are targeted towards self-tuned storage mechanism which must distribute content in the network to satisfy demands that vary
over time.

$\bullet$ \textbf{Analytical model:} We propose an analytical model 
to evaluate ICNs inspired by reliability theory concepts.  The model yields metrics such as the mean time to find a content and the 
load at the publishing areas as a function of the network topology and the time to live (TTL) of requests inside  domains.
Our model allows us to study tradeoffs in the choice of the TTL.  Smaller values of  TTL  
might lead to a reduction in the time to retrieve information at the cost of an increase in the system's load at publishing areas.

The remainder of this paper is organized as follows.  Section~\ref{related} presents related work and a background on 
 Information Centric Networks.  Section~\ref{system} introduces 
the proposed ICN system design, Section~\ref{analysis} contains the analytical
model and in Section~\ref{numerical}
 we present numerical results obtained with the proposed 
model.  Section~\ref{concl} concludes the paper.

%% file: literature.tex
\section{Related Work} \label{related}

The literature on ICN can be broadly classified into structured architectures, such as ~\cite{dona, psirp, netinf} and unstructured architectures, such as ~\cite{elisha1, bc+}. In what follows, we briefly describe some of the proposals.

Dona~\cite{dona} consists of an hierarchy of domains, wherein a resolution handler (RH) knows  
the IP location of all content published in descendent domains.    
 RHs placed in the highest domain are aware of all the content published in the entire network. 
Psirp~\cite{psirp}, Netinf~\cite{netinf} and Mulitcache~\cite{multicache} are proposals that deploy Distributed Hash Tables (DHT), 
enabling multiple logical entities to share the knowledge of all content published in the network, in a distributed fashion.
Each content placed in the network is mapped to a hash key, with a hash key being associated with a single node of the DHT.

DHT structures are highly scalable, but may impose considerable overhead maintenance costs~\cite{dhtmanagement}. 
Whenever a node fails or becomes repaired, requests must be sent along the network to reassure the node location in the network. 
Neighborhood adjacencies must be reestablished and file-index databases 
related to a partition management must be resettled. Also, many unsolved security vulnerabilities are able to disrupt the pre-defined 
operation of DHTs nodes ~\cite{securitydht}. At last, in a network composed of domains where providers 
care about administrative autonomy, 
the use of a global hash table is unfeasible~\cite{mdht}.

Jacobson et al.~\cite{ccn1} propose a non structured geometry topology for routing requests.
Published content is announced through routing protocols, composing routing tables supporting name aggregation.
Requests are routed towards publishing areas leaving a trail called \emph{bread crumbs}, 
so content follows the reverse paths 
set by the trails, when sent to users. As content flows to users, the bread crumbs are consumed. 
In general, content replicas will not be stored in the domain they were originally published. To be encountered, 
routes for replicas should be
in the routing tables. Avoiding an explosion of the size of the routing tables becomes an important challenge.

To enhance the discovery of cached contents, Rosensweig et al.~\cite{elisha1} and Kakida et al.~\cite{bc+} allow
 bread crumbs not to be consumed on the fly, 
when content traverses the network. This allows trails for previously downloaded contents to be preserved. 
A new user request that reaches a trail for a desired content will be sent to a cache-router indicated by the trail, before flowing to the publishing areas. 
Opportunistically downloading content from nearby cache-routers may considerably  reduce the time users take to retrieve information. 
Nevertheless, 
the proposals encompassing the preservation of bread crumbs do not explicitly address how published information will be 
announced to fulfill the routing tables.

In this article, we avoid the drawbacks mentioned in this section, for both
structured and unstructured geometry topologies. 
To this aim, we propose a novel architecture with cache-routers disposed across hierarchical domains. 
Users' requests flow across tiers towards the publishing areas. Random walks are issued to 
explore domains' vicinity, so as to allow opportunistic encounters with the desired content. 
That way, we avoid the drawbacks of structured geometries, as well as the problem of fulfilling routing tables which is common to  
the unstructured architectures discussed above.  

The model presented in this paper is inspired by reliability 
metrics~\cite{schoolrio1992}.  The performance of random walks 
for search in unstructured hybrid peer-to-peer systems has been analyzed by Ioannidis and Marbach~\cite{ioannidis2008design}.   Nonetheless, 
 Ioannidis and Marbach focus on asymptotic results when the number of nodes in the network grows to infinity, and considered a single
 domain (tier).  To the best of our knowledge, our work 
is the first to analyze the performance of random walks for search in multi-tiered architectures, accounting for the tradeoffs involved
 in the choice of the time to live of the random walks. Existing multicache multi-tier normal cache overlay network either consider global
knowledge of content placement (CDN, P2P), or single path to the storage area (Cache Trees). None of these methods explore the vicinity of 
a cache in a fully distributed way.

%% file: system.tex
\section{System Design} \label{system}

\begin{figure}[h!]
\center
\includegraphics[scale=0.3]{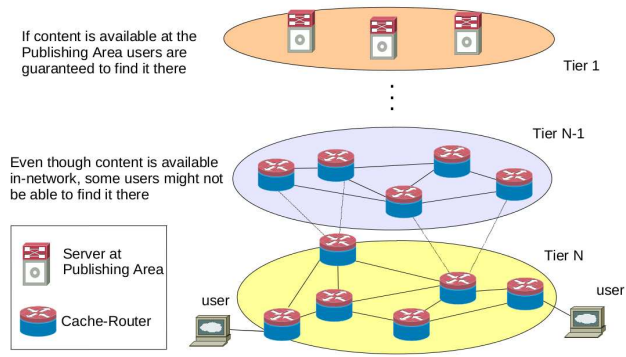}
\caption{ICN content distribution mechanisms.}
\label{greenred}
\end{figure}

Our architecture design considers \textbf{logical} hierarchical \emph{tiers} (also called \emph{domains})
composed of \emph{cache-routers} ($i.e.$, routers that can store content replicas).
We consider $N$ logical hierarchical tiers, in which tier 1 is the top level tier, and tier $N$ constitutes the bottom level.
Routers in tier $N$ are the only ones connected to users.
Users publish content in storage areas connected to tier $1$ routers.
The interaction among the publishing areas and tier 1 routers is better detailed in Figure~\ref{greenred}.
The figure displays a single publishing area in tier 1 and the logical hierarchy of cache-routers.
Routers forward  requests towards publishing areas which contain permanent copies of the content.
Replicas may be cached in the cache-routers in the logical hierarchy.
A strategy should be adopted to allow \emph{opportunistic} encounters between requests and replicas in a best-effort
manner as will be detailed below.
Opportunistic encounters are probabilistic in nature.

\subsection{Content search: random walks and bread crumbs}

When requests are issued by users the cache-routers forward them to the publishing areas.
Each cache-router of tier $i$ is logically connected to a subset of cache-routers in the same tier $i$
and in the parent tier $i-1$.
When a  request first reaches a tier, a random walk is issued to find replicas of the content that may have been
cached in that tier.  
The random walk lasts for at most $T$ units of time, and only traverses cache-routers in the same level.
That is, a counter is set to limit the amount of search time for a content {\em within a level}.
If the desired content is not found when this counter expires,
the router (say router $k$ at level $i$) that holds the request at that time transfers it to level $i-1$, that is,
to one of the routers in level $i-1$ router $k$ is (logically) connected to.
A new random walk will then be issued, but now at level $i-1$.
If the content been searched is found at a cache-router of hierarchy $j$, it is sent to the user as it will be explained below.
Otherwise, the request will be forwarded up in the hierarchy until the publishing area (tier 1) is reached where the content is
guaranteed to be found (otherwise the content was never stored in the first place). Note that the search does not generate 
significant traffic. Each request from a user generates a single search message that performs a random walk at one tier at a time.
The traffic due to this request message is negligible as compared to that generated by the content to be transferred.

As the requests are forwarded upwards across the hierarchy, a set of {\emph backward pointers} is maintained.
Those pointers are henceforth referred to as \emph{bread crumbs}.
When a content is located in the network, it is delivered to the users following the reverse path of the
bread crumbs. As the content follows the path of \emph{bread crumbs}, the trail is erased. Random walks within domains
do not generate bread crumbs. Therefore, when content is delivered to users it does not follow the exact same path as the 
random walks. This means that random walks do not impose extra hops/side effects when content is delivered to users. 
We use a set of counters, named \emph{reinforced counters} (RCs), to keep track of the load across the network for each content.
The placement of replicas will be controlled using the reinforced counters, 
which will be used to decide which and when contents should be stored or evicted at each cache-router. Our approach is
self-adaptive to users' loads.

\subsection{Content placement: reinforced counters}

Each request carries a hash key \textsf{\emph{inf}} to identify the content being searched for. 
As mentioned above, a trail of bread crumbs is left at each cache-router traversed by the request.
The trail includes a back pointer to the preceding router visited by the request (the bread crumb), 
the hash key \textsf{\emph{inf}} to identify the request and an associated counter
(called \emph{reinforced counter}) that is incremented when the request arrives at a cache-router.
The \emph{reinforced counter} is used to determine whether the content associated with \textsf{inf} 
should be stored at the cache-router. Each cache-router keeps a reinforced counter \textsf{rc(\emph{inf})}, 
associated to each hash key \textsf{\emph{inf}}. 
Whenever a request for \textsf{\emph{inf}} reaches a cache-router, \textsf{rc(\emph{inf})} is incremented by one.
When content is downloaded through the reverse path left by the trails,
bread crumbs are consumed but the reinforced counters are kept intact.
Each cache-router periodically decreases the reinforced counter \textsf{rc(\emph{inf})} associated to \textsf{\emph{inf}} and this
period is a parameter to be set.

A reinforced counter \textsf{rc(\emph{inf})} for content \textsf{\emph{inf}} at a cache-router has two thresholds: \textsf{rc-{up}(\emph{inf})} and \textsf{rc-{low}(\emph{inf})}.
If \textsf{rc(\emph{inf}) } reaches the upper threshold \textsf{rc-{up}(\emph{inf}) } the content \textsf{\emph{inf}} is cached at that cache-router
after it is found and when it traverses the cache-router towards to the user.
Whenever \textsf{rc(\emph{inf})} reaches the lower threshold \textsf{rc-{low}(\emph{inf})} and if
\textsf{\emph{inf}} is cached at that cache-router, this content is immediately removed from the cache-router. 
We assume that cache-routers have enough storage capacity to store any content that they are required to maintain. 
The reinforced counters mechanism allows popular contents to be transferred from the publishing areas to the cache-routers.  
That way, the opportunistic discovery of popular content is favored at cache-routers.  
In contrast, if demand is not high enough to justify consumption of storage resources, reinforced
counters will favor the eviction of the content. 

We provide additional insights on the connection between reinforced counters and content placement. For simplicity of exposition, and without loss of generality, we assume
that content is demanded by users at a given fixed rate.
We then show that the reinforced counters fully determine content placement at the cache-routers. 
To this aim, we consider a fluid approximation, where content requests arrive according to a flow with a given intensity,
and the reinforced counters are also approximated as being continuous. Let $\Lambda_{m,l}(\textsf{\emph{inf}})$ be the total load for \textsf{\emph{inf}} arriving from tier $l+1$ at the $m$-th cache-router at tier $l$. 
Let $\gamma(\textsf{\emph{inf}})$ be the rate at which the reinforced counters \textsf{rc(\emph{inf})} are decremented.   
We assume that  the \textsf{rc(\emph{inf})} are initialized as \textsf{rc-low(\emph{inf})}, so that if 
 $\Lambda_{m,l}(\textsf{\emph{inf}}) = \gamma(\textsf{\emph{inf}})$ then content will not be stored at $m$.

\vspace{0.4in}
\begin{proposition} Under the fluid approximation and an hierarchical tiered topology, a cache-router stores a copy of \textsf{\emph{inf}} if and only if $\Lambda_{m,l}(\textsf{\emph{inf}}) > \gamma(\textsf{\emph{inf}})$. \label{propofixedload}
\end{proposition}
\vspace{-0.1in}
{\bf Proof:}
The proof is inspired by~\cite[Lemma 1]{breadcrumbs1}.  In an hierarchical network we define the direction of requests as \emph{upstream} and the reverse as \emph{downstream}.  A cache-router is affected only by requests that pass through it, and requests pass through routers
only upstream, so upstream cache-routers do not impact cache-router $m$ through their requests.
Content flows downstream, but only passes through $m$ for requests that were issued by cache-router $m$. 
Therefore, the state and requests of upstream routers do not impact the state and requests issued by downstream routers.  

We can determine if a cache-router will store a content in a bottom-up manner.  
First, consider the leafs of the tiered topology. 
At such cache-routers, if $\Lambda_{m,l}(\textsf{\emph{inf}}) > \gamma(\textsf{\emph{inf}})$ then \textsf{rc(\emph{inf})} will eventually reach \textsf{rc-up(\emph{inf})}, and content will be stored and never evicted.  
If $\Lambda_{m,l}(\textsf{\emph{inf}}) \leq \gamma(\textsf{\emph{inf}})$, in contrast, \textsf{rc(\emph{inf})} will remain equal to 
its initial value \textsf{rc-low(\emph{inf})}. \textsf{rc(\emph{inf})} will afterwards remain fixed,
and the cache-router, marked. Once  leafs are marked,  we proceed upstream. 
Using the same argument as the one in the paragraph above, we determine the content placement at the next cache-router that has all its children marked, and so on, up to reaching routers at tier 1, where all content is stored.  \hfill $\Box$ 

Given a workload, proposition~\ref{propofixedload} can be used to determine which domains across the network will hold a copy of
each content, if we know the (probabilistic) choice by which a router of level $i$ passes requests to routers of level $i-1$
at which it is logically connected to.
In the remainder of this paper we will assume that the probability distribution of content over cache-routers
is fixed and given for all cache-routers.

%% file: analysis.tex
\section{System Analysis} \label{analysis}

\begin{figure}[h!]
\center
\begin{tabular}{cc}
\vspace{-0.1in}
\includegraphics[scale=0.80]{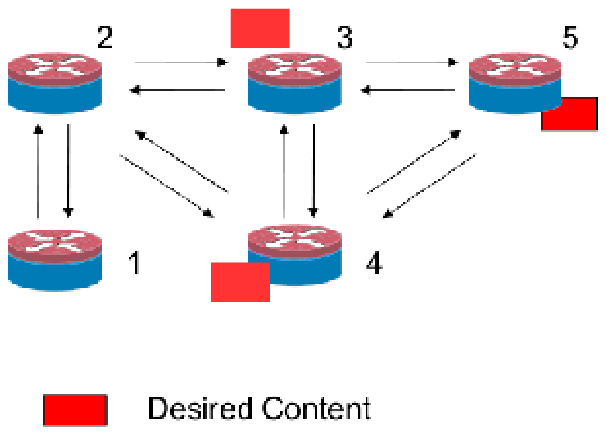} & \includegraphics[scale=0.35]{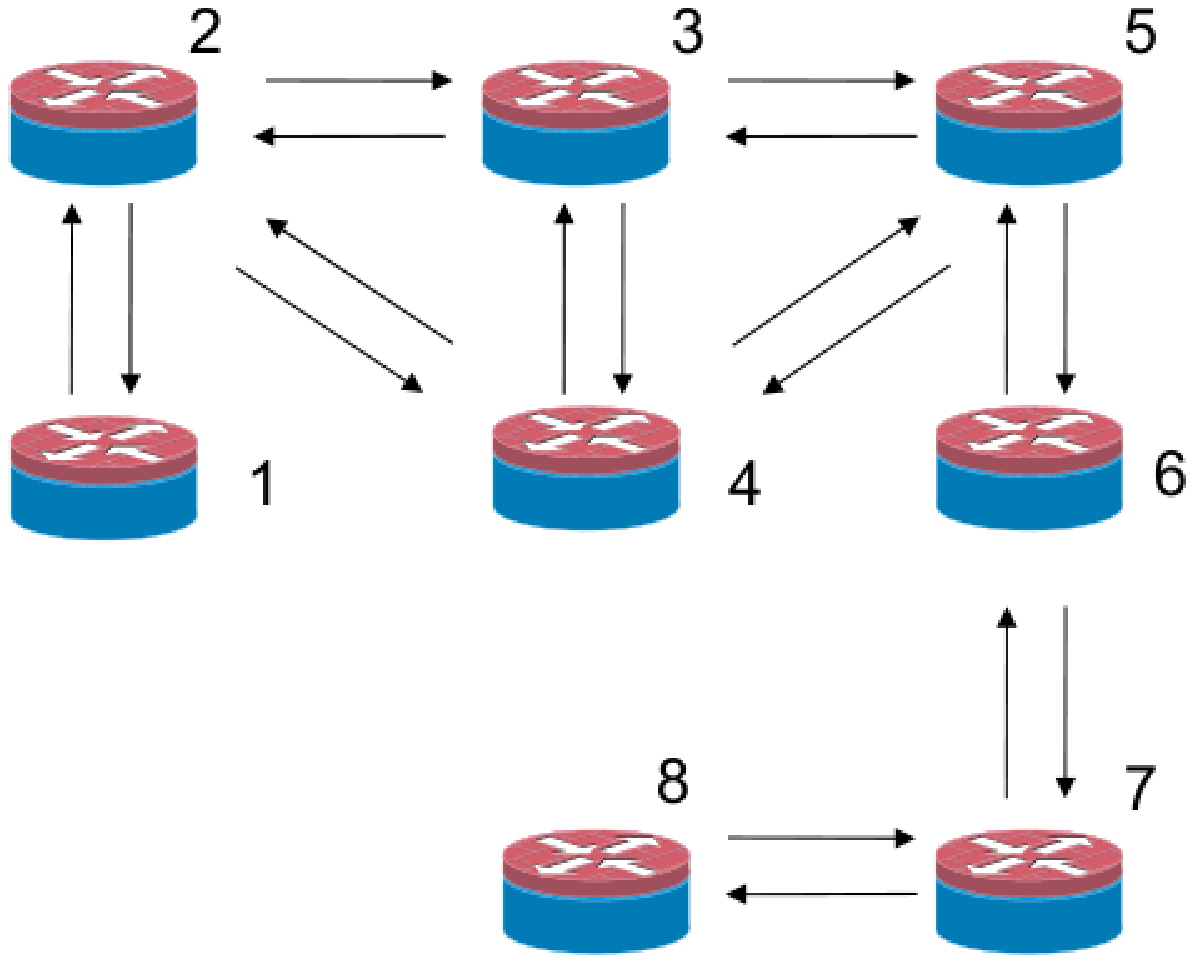} \\
\vspace{-0.1in}
(a) & (b) \\
\vspace{-0.1in}
\end{tabular}
\caption{(a) Example of a domain with five cache-routers and 
(b) Example of a domain with eight cache-routers.}
\label{modified_pm}
\end{figure}

In this section we present the model analysis for a single domain (i.e., a logical tier).  
Our main goals are to derive the key metrics of interest, namely 
1) mean time to find content and 
2) the fraction of requests that hit the publishing area.
These metrics are important to evaluate the performance of the proposed architecture. 
We consider a network of $C$ cache-routers in a domain. 
Figure~\ref{modified_pm}(a) illustrates a domain with five cache-routers. 
Assume that the desired content is stored at cache-routers 3, 4 and 5.
Suppose that a request arrives from a downstream domain to cache-router 2.
Then, cache-router 2 starts a random walk in the domain to find the content.  
Let $T$ be the maximum time a request might
spend in a domain before being redirected to the upstream domain in the hierarchy.  
$T$ is also referred to as time to live, or TTL. 

We consider a continuous time random walk in which the time between walker movements
are exponentially distributed with rate $\psi$, with the associated infinitesimal
generating matrix $Q$. 
Let $\Delta(i)$ be the out-degree of node $i$ and $\psi$ the walker rate. 
Then, all diagonal elements in $Q$ are equal to $-\psi$, and  $q_{ij}=\psi/\Delta(i)$
if there is a logical link from $i$ to $j$.   
We use uniformization ~\cite{unifsurv01} to obtain the metrics of interest.  
Let  $P$  be  the  \emph{uniformized  matrix}, obtained  from  $Q$  as
$P=I+{Q}/{\Lambda}$, where $\Lambda$ is a positive number greater than
the maximum  absolute value  of the elements  in the diagonal  of $Q$.
$\Lambda$   is   also   referred   to  as   the   
\emph{uniformization rate}.   

\begin{table}
\center
\begin{tabular}{l|l}
\hline
variable & description \\
\hline
\hline
$C$ & number of cache-routers in the network \\
$T$ & time to live \\
$H(T)$ &  time to hit content \\
$L(T)$ & lifetime of random walk in a domain \\
$R(T)$ & probability of not hitting content by $T$ \\
$T_0$ & time cost for accessing the publishing area \\
$p$ & probability that content is in domain \\
\hline
\end{tabular}
\caption{Table of notation. Time to live, $T$, is integer when considering discrete 
time random walks, and real when considering continuous time random walks.}
\label{notation}
\end{table}

Let \begin{math} \Omega_P(k)\end{math}  be the probability that, after
$k$ transitions  of the uniformized  process, the random walk  had not
hit  the desired  content. 
To compute  \begin{math} \Omega_P(k) \end{math}  we construct  a modified
transition matrix $\widetilde{P}$, assuming that the desired content is
placed in a subset of cache-routers of the domain.
Let $\omegavec$ be a column vector of size $C$ where 
$\omegavec_{i}=p_{i}$ and $p_{i}$ is the probability that cache-router $i$ stores the 
content, $\sum_{i=1}^{C} p_{i}=1$. 
To facilitate the notation, we number the cache-routers where $p_i > 0$ 
with the highest indexes of matrix $P$.
In Figure~\ref{modified_pm}(a) they are numbered 3,4 and 5. 
Then, $\widetilde{P}$ is defined as
\begin{equation}
\widetilde{P}=P (I - (diag(\omegavec)))
\label{eq:P-til}
\end{equation}
where $diag(\omegavec)$ is a matrix with the diagonal elements equal to 
vector $\omegavec$ and all other elements equal to zero. 
$\widetilde{P}$ is a sub-stochastic matrix where 
the sum of the elements in line $i$ is the probability of not finding 
the content in the domain, in one step, given that the random walk starts at cache-router $i$.
The element $(i,j)$ of $\widetilde{P}$ is the probability that a random 
walker that starts at cache $i$ moves to cache $j$ in one step and then a cache miss occurs at
$j$. 

Let  $\pi(0)$ be  the initial  state probability  vector indicating 
from where the search starts in the domain.
The $i$-th element  of $\pi(0)$, $\pi_i(0)$, corresponds  to the  probability that
the random walk  starts at cache-router $i$.  
The $(i,j)$ entry of matrix $\widetilde{P}^k$ characterizes the probability of visiting
state $j$ after jump $k$, given that the system  starts at state $i$.
Let  $\upsilonvec_{\widetilde{P}}(k)$ be a vector whose $m$-entry is 
the probability  that after $k$ steps  a  miss occurs at cache-router  $m$, given that the
random walk initial state is  $\pi(0)$. The following recursion can be used to compute $\upsilonvec_{\widetilde{P}}(k)$,
\begin{equation} 
\upsilonvec_{\widetilde{P}}(k) = \upsilonvec_{\widetilde{P}}(k-1) \widetilde{P}, \qquad k > 0
\label{eq:upsilon} 
\end{equation}
where $\upsilonvec_{\widetilde{P}}(0)=\pi(0)$. 
Then, $\Omega_P(k)$ is given by
\begin{equation}
\Omega_P(k) =\upsilonvec_{\widetilde{P}}(k) {\bf U}, \qquad k \ge 0
\label{eq:omega}
\end{equation}
where ${\bf U}$ is a column vector with all elements equal to one.     

\subsection*{Mean Time to Reach Content}

Recall that $T$ is the maximum time to live of a random walk in a domain,  $T \in \mathbb{R}$.
Let $R(T)$ be the probability that the content is not found by $T$ units of time. 
Then, conditioning on the number of transitions in the interval $[0,T]$, 
\begin{equation}  
R(T) =  \sum_{n=0}^{\infty} \varphi(n,T)\Omega_P(n) 
\label{rtini} 
\end{equation}
where $\varphi(n,T) = \exp({-\Lambda  T}) (\Lambda  {T})^n/{n!}$ is the probability  
that $n$ jumps occur  in the
interval $[0,T]$.  The number of jumps in the interval $[0,T]$ is Poisson distributed as we uniformized $Q$.

Next, our goal  is to compute the expected hitting  time of a content,
$E[H(T)]$. 
The  expected hitting time of  a content is the  sum of two
components.  
The  first component characterizes  the mean time  to hit
the content given that it is available in the domain, while the second
component  characterizes the mean  time to  hit the  content if  it is
unavailable in  the domain.  
In case  the content is  available in the
domain, the \emph{lifetime of the  random walk} in the domain (i.e., the searching
time in the domain), $L(T)$,
is the  minimum between the time  until reaching the  content and $T$. The mean of $L(T)$ is  given by
\begin{equation}
E[L(T)]=\int_0^T R(t)dt
\end{equation}
where $\lim_{T\rightarrow \infty}E[L(T)]$ is referred to as 
\emph{mean time to exit} in the reliability literature, 
and can be evaluated using standard techniques ~\cite{transol2000}.   

Let  $p$ be  the  probability that  the  content is  available in  the
considered domain.  If the  content is available, the probability that
the content  is not found  by time $T$  is $R(T)$.  If the  content is
unavailable, the random walk will take time $T$ inside the domain, and
additional $T_0$  units of time  to reach the  publishing area.
Therefore, the expected hitting time of a content is
\begin{equation} 
E[H(T)] = \left(E[L(T)]+  T_0R(T)\right)p +(T_0+T)(1-p) 
\label{ehtal} 
\end{equation}

Replacing~\eqref{rtini}  into~\eqref{ehtal}, and exchanging  the order
of the integration and summation, yields, after simplification,
\begin{equation} 
E[H(T)] = \left(\sum_{n=0}^{\infty} \left(1-\sum_{m=0}^{n} e^{-\Lambda T} \frac{(\Lambda T)^m}{m!} \right)  
\Lambda^{-1} \Omega_P(n)  +  T_0 R(T)\right)p +(T_0+T)(1-p) 
\label{eqeht} 
\end{equation}

The analysis above easily handles discrete time random walks, where
time between walkers movement is fixed.  

%% file: experiments.tex
\section{Experimental Results and System Analysis}
\label{numerical}

In this section we present numerical results for a few examples in order to
illustrate our proposal and the existing tradeoffs in the choice of parameter values.
In the development of section \ref{system},
the time intervals between the random walker steps can 
be either independent and exponentially distributed random variables or constant. 
For the numerical studies we consider constant step values.

Our goals are to
1) show the impact of different system parameters on the mean time
to find content and
2) evaluate the average load on publishing area servers, which impacts both the
time to recover a content and the system's scalability.
We analyze a simple two-tier setup wherein the publishing area servers have finite capacity.
This simple architecture suffices to highlight the key points we want to
emphasize in this work while keeping the model description short. 
From this scenario we also indicate how the evaluation could easily consider 
multiple logical tiers.

Our reference topology is that illustrated in Figure~\ref{modified_pm}(a).
In this topology, there is a single content that may be cached in one of the cache-routers $3, 4$ or $5$ with
equal probability of residing in any of these three routers, given it is in this logical tier.
Therefore, $p_u=1/3$ for $u=3,4,5$.
With probability $1-p=0.5$, the content does not reside in the tier and it takes and additional 
$T_0=100$ time units to find it in the publishing area.
The parameter $T$, the maximum time the random walker is allowed to search for the
content in a logical tier is varied between 1 and 200.
The parameters values may vary according to the experimental goals.

\subsection{Single Domain Case}
\label{singledomain}

\begin{figure}[h!]
\center
\begin{tabular}{c@{\hskip 0.1in}c}
\hspace{-0.4in}
\includegraphics[scale=0.50]{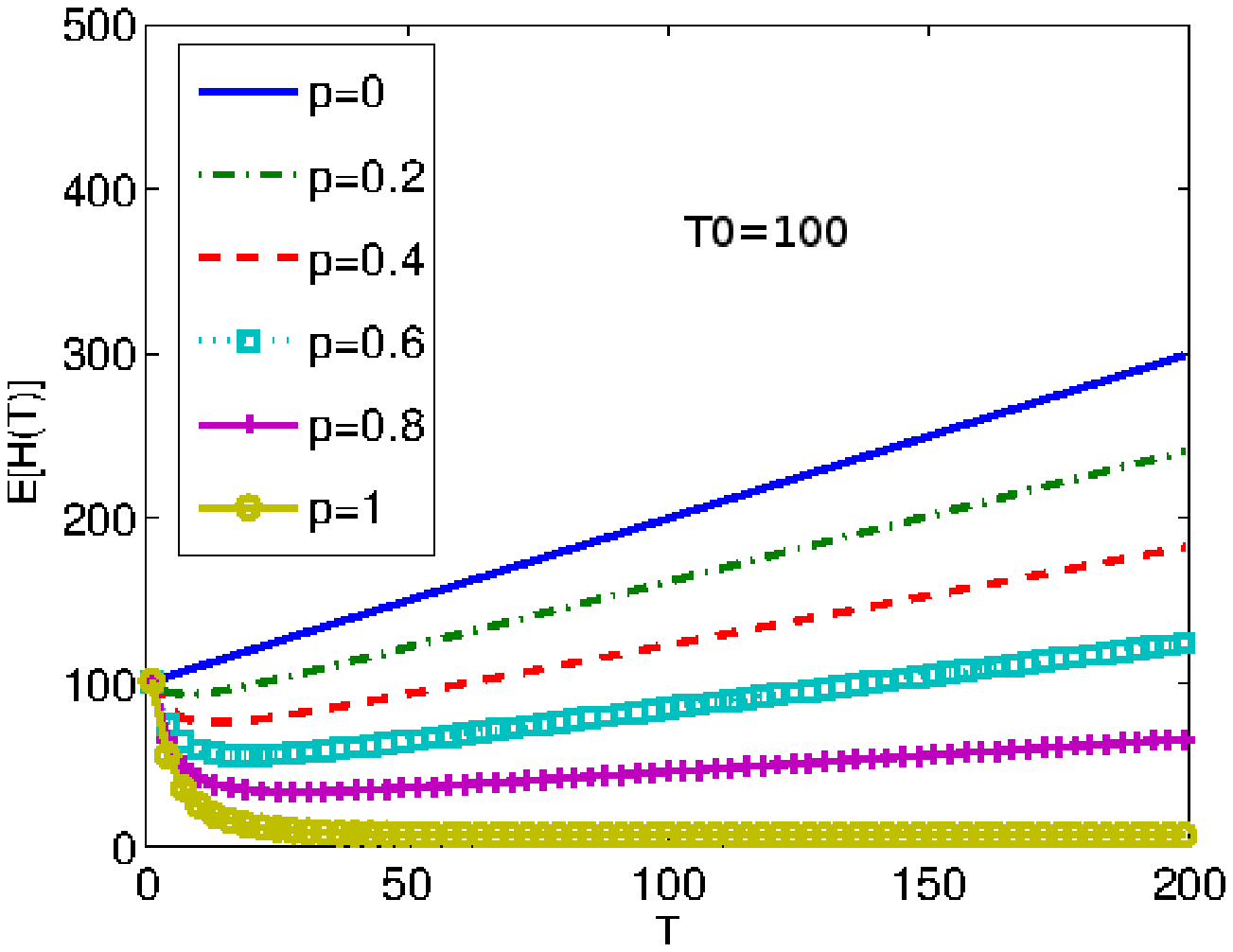} & \includegraphics[scale=0.50]{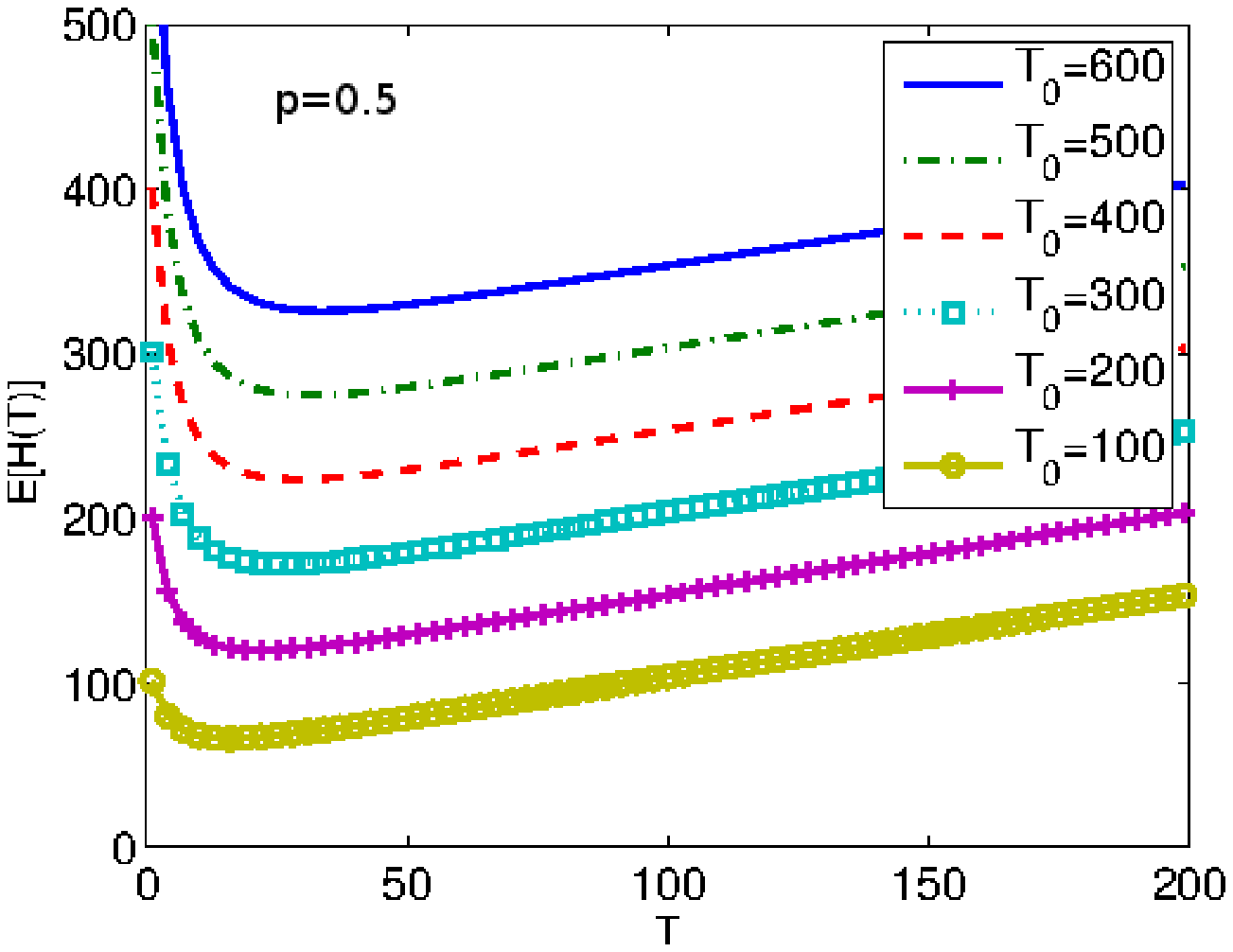} \\
\vspace{-0.1in}
(a) & (b) 
\end{tabular}
\caption{Effect of $p$ and $T_0$ on the expected hitting time}
\label{optimal_p} \label{optimal_t0}
\end{figure}

We illustrate how the mean time to find the content, $E[H(T)]$, depends on different system parameters.  
Figure~\ref{optimal_p}(a) shows $E[H(T)]$, obtained using~\eqref{ehtal}, as a function of $T$,
for different values of $p$ varying between 0 and 1, with increments of 0.2.
The topology and the content placement are kept fixed.  
Let  $T^{\star}$ be the optimal value of $T$ that minimizes $E[H(T)]$. 
As the probability that the content is available increases, $T^{\star}$ increases.
Clearly, when $p=0$, $T^{\star} = 0$, since if the content is not available in the tier it is better
not to search for it. 
On the other hand, when the content is always available ($p=1$), $T^{\star} \rightarrow \infty$.
This is because, in this example, $T_0$ is large compared to the mean time to find the content
in the tier ($E[H(\infty)]$). 
In this scenario, it will take less time, on average,  for the random walker to find the content in the domain, 
as opposed to find it in the publishing area servers, if the walker is given enough time to search for the content in
the domain (i.e., if $T$ is very large).
For values of $p$ between $(0,1)$, the optimum value $T^{\star}$ can be obtained from our model
as shown in Figure \ref{optimal_p}(a).

For any given value of $p$, Figure~\ref{optimal_p}(a) shows that $E[H(T)]$ first decreases and then increases, 
except for $p=0$ and $p=1$, when $E[H(T)]$  always increases and decreases, respectively.
In any case, $E[H(T)]$ asymptotically grows at rate $1-p$, as can be inferred from equation \eqref{ehtal}. 
The impact of $T_0$ (the time cost for accessing the publishing area servers)
on $E[H(T)]$ is illustrated in Figure~\ref{optimal_t0}(b),
which shows $E[H(T)]$ as a function of $T$, for $T_0 \in [100, 200, \ldots, 600 ]$.
As \begin{math} T_0 \end{math} decreases,
there is a subtle decrease of the optimal time to find a content in the domain.
This is because, as the time to access publishing area servers increases,
it is beneficial to remain longer in the domain before forwarding the request to other domains.

Until now we analyzed the problem of content discovery from the clients perspective.
The administrators of publishing areas, in turn, might worry about the traffic
that is incurred in their servers. 
If content is available in the cache-routers in a tier, the load from that domain to the servers
(assuming a single tier architecture) is proportional to $R(T)$,
the probability that users do not find the content in the tier by time $T$.  
In more detail, let ${\lambda}$ and $\Lambda$ be the average user load (number of requests for a content per time unit)
that arrives to a tier and the average load flowing from the tier to the next in the hierarchy
(or from the tier to the publishing area), respectively. 
The number of requests per unit time served inside the domain is clearly ${\lambda}-\Lambda$
and ${\Lambda}= R(T) \lambda$.
For the plots we consider $\lambda = 100$.

\begin{figure}[h!]
\center
\begin{tabular}{c@{\hskip 0.1in}c}
\hspace{-0.3in}
\includegraphics[scale=0.45]{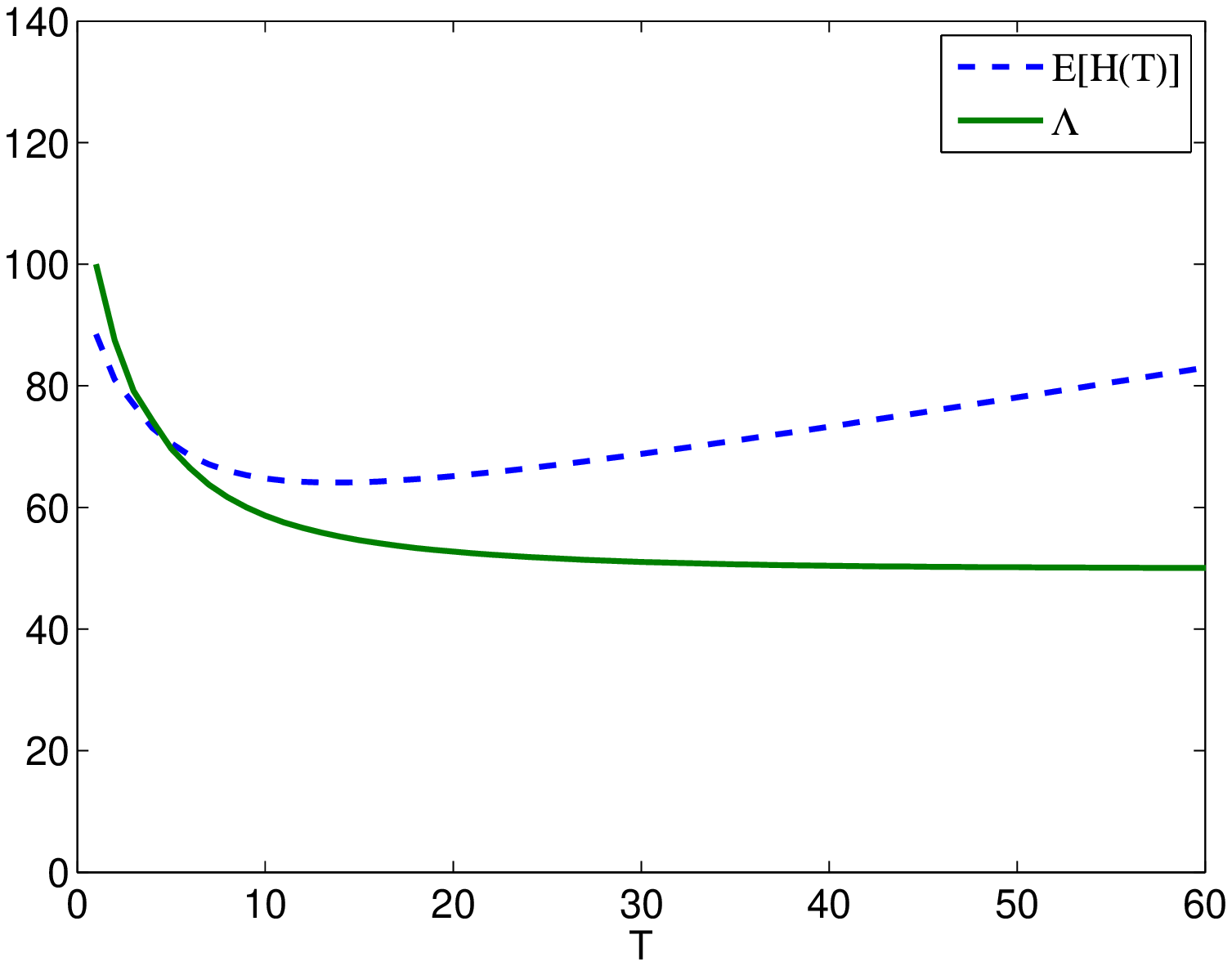} &
\includegraphics[scale=0.45]{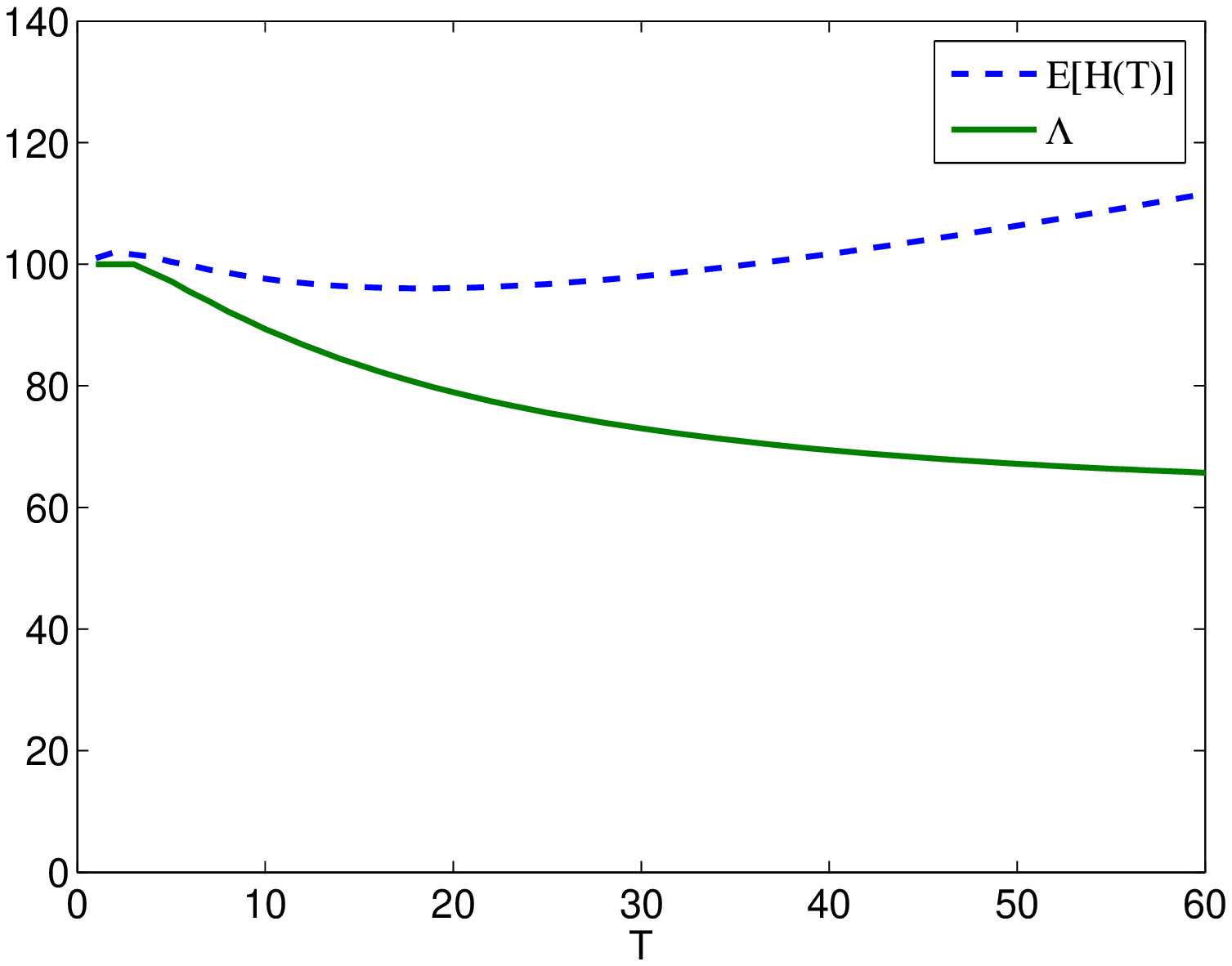}\\
\vspace{-0.1in}
(a) & (b) 
\end{tabular}
\caption{Probability of not hitting content by time $T$, $E[H(T)]$, and $\Lambda$, for (a) 5 cache topology, (b) 8 cache topology.}
\label{reliability_1}
\end{figure}

Figures~\ref{reliability_1}(a) and~\ref{reliability_1}(b) show how $\Lambda$ varies as a function of  $T$.   
Figures~\ref{reliability_1}(a) considers the reference scenario, modified so that content may be available at
caches 2, 3, 4 and 5 with probability 0.25 at each, given the desired content is in the domain.
Figures~\ref{reliability_1}(b), in turn, is generated from the topology with 8 caches
shown in Figure~\ref{modified_pm}(b), where content may be available at caches 5, 6, 7 and 8 with
probability 0.25 at each, given the desired content is in the domain.
The other parameters are the same as in the reference scenario. 

Figures~\ref{reliability_1}(a) and~\ref{reliability_1}(b) indicate that 
there is a tradeoff between users interests and providers time-cost when choosing the optimal value of $T$. 
If $T$ is selected so as to minimize $E[H(T)]$, the rate at which the publishing area servers are accessed,
$\Lambda$ might be considerably large.  
However, a slight increase in  $T$ and $E[H(T)]$ might yield significant reductions in $\Lambda$.
Therefore, we may minimize $E[H(T)]$ constraining $\Lambda$ to a given (relatively small) value,
that is, we may be able to keep the load at the publishing area servers at a low value
and yet avoiding a considerable increase in the time to retrieve the searched content.

Note that in Figure \ref{reliability_1}(b) $\Lambda$ is constant for $T \leq 3$. 
This is because we are considering a discrete time random walk for the studies. 
In this case, the probability that the content is found inside the domain is zero if $T$ is smaller
than the distance between the source of the random walker and the cache-router that may have the content. 
In a continuous time random walk, in turn, $R(T)$ is strictly decreasing with respect to $T$.

\setlength{\tabcolsep}{0.1pt}

We may summarize the above findings as follows.
If content is available in a tier with high probability, it may be advantageous to increase 
the time the random walker is allowed to search for a content in the tier to reduce the load
at the publishing area servers.
On the other hand, the expected time to find the content may increase.
Quantifying these tradeoffs is a contribution of our work.

\subsection{Multiple Domain Case}

\begin{figure}[h!]
\center
\begin{tabular}{c@{\hskip 0.1in}c}
\hspace{-0.2in}
 \includegraphics[scale=0.52]{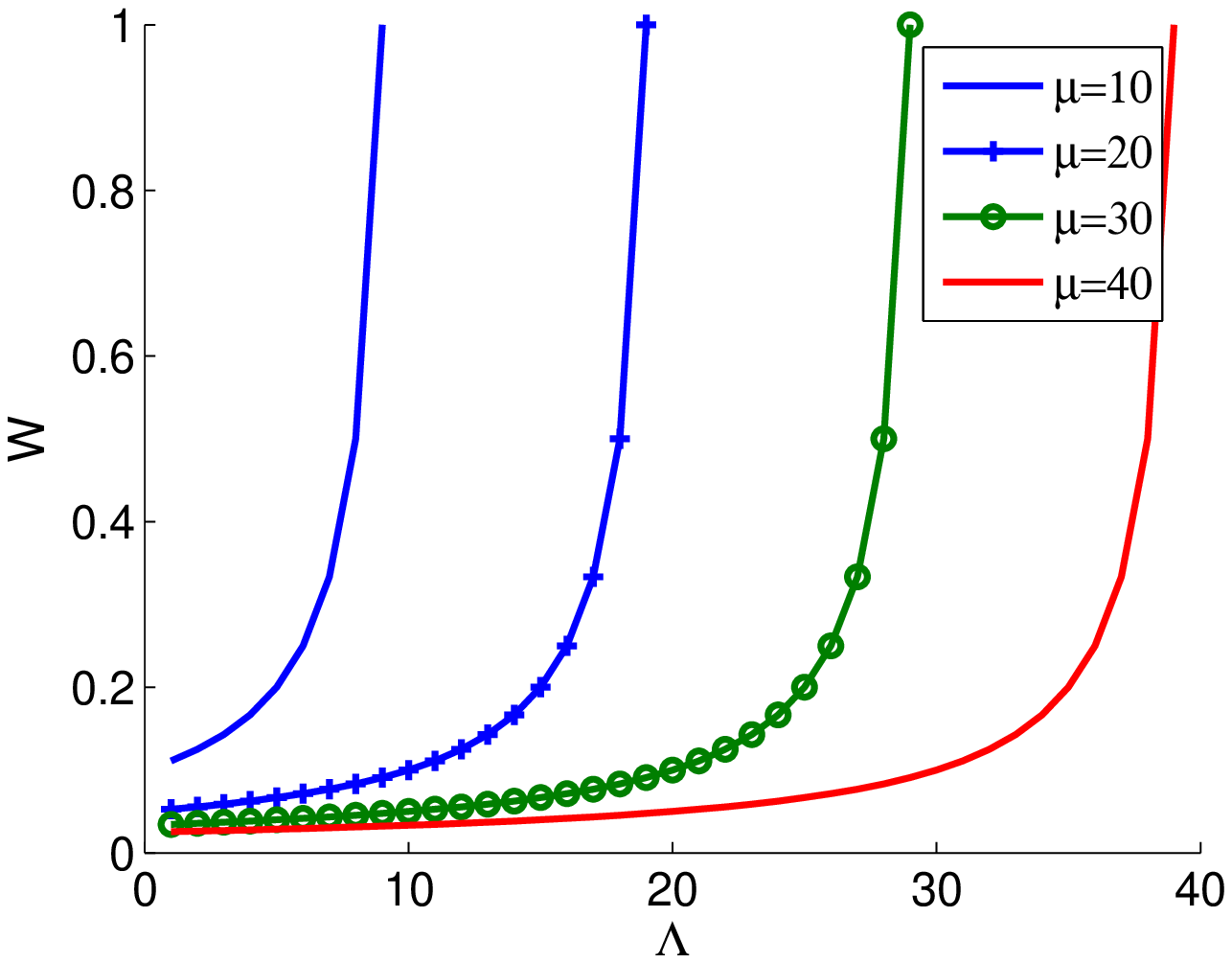}  & \includegraphics[scale=0.52]{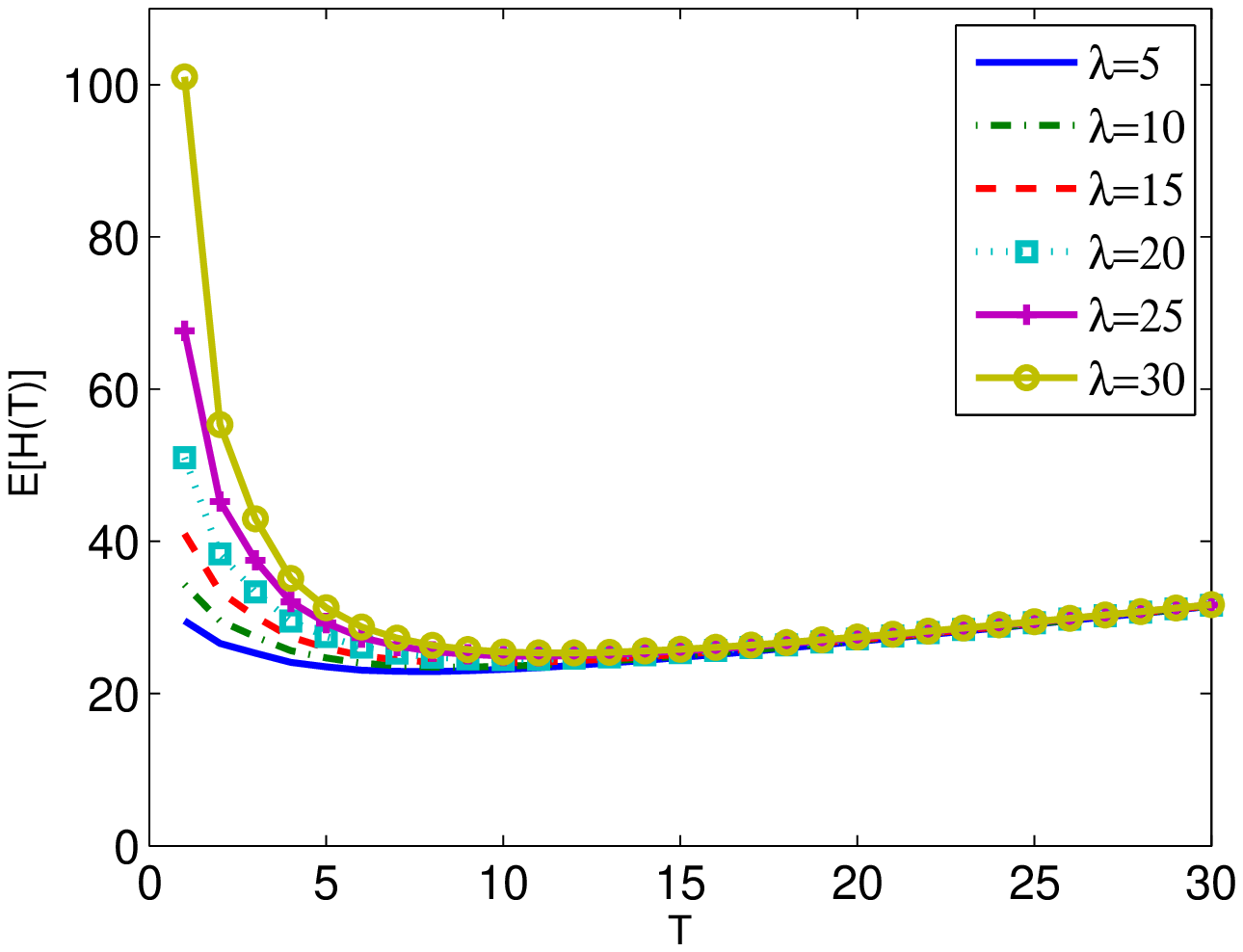} \\
(a)& (b) 
\end{tabular}
\caption{(a) Average server processing time as a function of $\Lambda$, (b) mean time to find content, accounting for server capacity $\mu=40$.}
\label{mm1}
\end{figure}

In the previous section we assumed a constant time $T_0$ to retrieve a content from the publishing area servers.
In what follows we consider a scenario where content that is not found in a
domain is forwarded to publishing area servers with limited capacity.
Therefore, the time cost to access the publishing area servers, $T_0$, will now be a function of the amount of 
traffic $\Lambda$ that is directed to the publishing area servers (which in turn is a function of $R(T)$). To illustrate 
the impact of the load of users on the publishing area servers,
let us a consider a simple M/M/1 model for the server, with associated service capacity $\mu$.
Then, the mean waiting time experienced by users accessing the server is  
$W=1/(\mu-{\Lambda})$.  
Figure~\ref{mm1}(a) shows how $W$ varies as a function of ${\Lambda}$. 
In what follows, for illustration purposes, we set $\mu=40$ and $T_0 = 1000 W$,
\begin{equation}
 T_0=1000/(\mu - R(T) \lambda )
\end{equation}
Figure~\ref{mm1}(b) illustrates the impact of the service capacity on the mean waiting time experienced
by users and shows how $E[H(T)]$ varies as a function of $T$.  
The qualitative behavior of $E[H(T)]$ is the same as the one observed previously, and the insights obtained 
in Section~\ref{singledomain} about the tradeoff in the choice of $T$ still apply.  
Note that an increase in $R(T)$, the probability of not finding the searched content in the tier
(not shown in the figure) may degrade users performance, as it may lead to server overload. 
In particular, if $T$ is small (e.g., $T = 1$), $E[H(T)]$ grows unbounded with $\lambda$. Due to space limitations 
we were only able to discuss a few examples aiming at describing our proposal and
the existing design tradeoffs. Our approach, however, can handle multiple domains and user loads as we briefly sketch below.
From the results of section \ref{analysis} and given the user loads, we can obtain the
domains that hold copies of the content.
$R(T)$ is then used to obtain the loads at each domain as well as at the publisher. Finally, given the loads at the different domains, $E[H(T)]$ is obtained in a top down fashion,
starting from the top and moving to the lowest level domains. 
%
%

%% file: conclusion.tex
\section{Conclusion}
\label{concl}

Information centric networks are gaining considerable attention from researchers and practitioners
due to their scalability and robustness properties. 
Nonetheless, there are still important challenges in the design, modeling and analysis of such networks. 
From the design perspective, one of the challenges is to determine how to fill out the routing tables. 
From the modeling and analysis perspective, the challenges are associated to the large number of routers envisioned in ICNs. 
In this paper, we have proposed a novel design for ICNs that is built on top of
random walks and hierarchical tiers (domains) to cope with the exploration \emph{versus}
exploitation tradeoff involved in content search. 

Our model computes metrics such as mean time to
find a content and evaluate existing tradeoffs to tune the parameters of the
architecture we propose. In this present work we focus on the performance tradeoffs of the architecture.
However, our analytical model can be extended to study the scalability of the proposed architecture
with respect to others in the literature. 
For instance, as the user load for a content grows what is the overall increase in the
expected time to find a content as compared to a P2P architecture.
This work also opens additional avenues for future research.  
One such problem is to determine the impact of parallel 
random walks across tiers at the same level of the hierarchy as well as  across
different levels.

%% file: cache.bbl
\begin{thebibliography}{}

\bibitem[2nd NetInf Description 2010]{netinf}
2nd NetInf Description (2010).
\newblock Netinf.
\newblock \url{http://www.4ward-project.eu/}.

\bibitem[Ahlgren et~al. 2012]{dirkkutscher}
Ahlgren, B., Dannewitz, C., Imbrenda, C., Kutscher, D., and Ohlman, B. (2012).
\newblock A survey of information-centric networking.
\newblock {\em IEEE Communications Magazine}, 50(7):26--36.

\bibitem[Chen et~al. 2008]{dhtmanagement}
Chen, S., Zhang, Z., Chen, S., and Shi, B. (2008).
\newblock Efficient file search in non {DHT} {P2P} networks.
\newblock {\em Elsevier Computer Communications}, 31(2):304--317.

\bibitem[D'Ambrosio et~al. 2011]{mdht}
D'Ambrosio, M., Dannewitz, C., Karl, H., and Vercellone, V. (2011).
\newblock {MDHT}: A hierarchical name resolution service for
  information-centric networks.
\newblock In {\em Proc. of SIGCOMM workshop on ICN}, pages 7--12.

\bibitem[de~Souza~e Silva and Gail 2000]{transol2000}
de~Souza~e Silva, E. and Gail, H.~R. (2000).
\newblock {Transient Solutions for Markov Chains}.
\newblock In Grassmann, W., editor, {\em Computational Probability}, pages
  44--79. Kluwer.

\bibitem[de~Souza~e Silva and Gail 2001]{unifsurv01}
de~Souza~e Silva, E. and Gail, H.~R. (2001).
\newblock {The Uniformization Method in Performability Analysis}.
\newblock In B.~R.~Haverkort, R.~Marie, G.~R. and Trivedi, K.~S., editors, {\em
  Performability Modelling: Techniques and Tools}, chapter~3, pages 31--58.
  Wiley.

\bibitem[de~Souza~e Silva et~al. 2006]{rio06}
de~Souza~e Silva, E., Leão, R.~M.~M., Santos, A.~D., Azevedo, J.~A., and
  Netto, B.~C.~M. ({2006}).
\newblock {Multimedia Supporting Tools for the CEDERJ Distance Learning
  Initiative applied to the Computer Systems Course}.
\newblock In {\em {22th ICDE World Conference on Distance Education}}, pages
  {1--11}.

\bibitem[de~Souza~e Silva and Muntz 1992]{schoolrio1992}
de~Souza~e Silva, E. and Muntz, R.~R. (1992).
\newblock {\em {M\'etodos Computacionais de Solu\c{c}\~ao de Cadeias de Markov:
  Aplica\c{c}\~oes a Sistemas de Computa\c{c}\~ao e Comunica\c{c}\~ao}}.
\newblock SBC - Escola de Computa\c{c}\~ao.

\bibitem[Ghodsi et~al. 2011]{scott}
Ghodsi, A., Schenker, S., Koponen, T., Singla, A., Raghavan, B., and Wilcox, J.
  (2011).
\newblock Information-centric networking: Seeing the forest for the trees.
\newblock In {\em Proc. of HOTNETS}, pages 1--6.

\bibitem[Ioannidis and Marbach 2008]{ioannidis2008design}
Ioannidis, S. and Marbach, P. (2008).
\newblock On the design of hybrid peer-to-peer systems.
\newblock {\em ACM SIGMETRICS Performance Evaluation Review}, 36(1):157--168.

\bibitem[Jacobson et~al. 2009]{ccn1}
Jacobson, V., Smetters, D.~K., Thornton, J.~D., Plass, M.~F., Briggs, N.~H.,
  and Braynard, R.~L. (2009).
\newblock Networking named content.
\newblock In {\em Proc. of CoNEXT}, pages 1--12.

\bibitem[Kakida et~al. 2011]{bc+}
Kakida, M., Tanigawa, Y., and Tode, H. (2011).
\newblock Breadcrumbs+ : Some extensions of naive breadcrumbs for in-network
  guidance in content centric networks.
\newblock In {\em Proc. of IEEE IPSJ}, pages 376--381.

\bibitem[Katsaros et~al. 2011]{multicache}
Katsaros, K., Xylomenos, G., and Polyzos, G.~C. (2011).
\newblock Multicache: An overlay architecture for information-centric
  networking.
\newblock {\em Elsevier Computer Networks}, 55(4):936--947.

\bibitem[Koponen et~al. 2007]{dona}
Koponen, T., Chawla, M., Chun, B.~G., Ermolinskiy, A., Kim, K.~H., Shenker, S.,
  and Stoica, I. (2007).
\newblock A data-oriented (and beyond) network architecture.
\newblock In {\em Proc. of SIGCOMM}, pages 181--192.

\bibitem[Lagutin et~al. 2010]{psirp}
Lagutin, D., Visala, K., and Tarkoma, S. (2010).
\newblock Publish/subscribe for internet: {PSIRP} perspective.
\newblock In {\em Emerging Trends from European Research, (Valencia FIA book
  2010)}.

\bibitem[Menasché et~al. 2010]{selfsust}
Menasché, D.~S., Rocha, A.~A., de~Souza~e Silva, E., Leão, R.~M.~M., Towsley,
  D., and Venkataramani, A. (2010).
\newblock Estimating self-sustainability in peer-to-peer swarming systems.
\newblock {\em Perf. Eval.}, 67(11):1243--1258.

\bibitem[Rosensweig et~al. 2010]{elisha1}
Rosensweig, E., Kurose, J., and Towsley, D. (2010).
\newblock Approximate models for general cache networks.
\newblock In {\em Proc. of INFOCOM}, pages 1--9.

\bibitem[Rosensweig et~al. 2013]{breadcrumbs1}
Rosensweig, E., Menasché, D., and Kurose, J. (2013).
\newblock On the steady-state of cache networks.
\newblock In {\em Proc. of INFOCOM, to appear}.

\bibitem[Urdaneta et~al. 2011]{securitydht}
Urdaneta, G., Pierre, G., and Steen, M.~V. (2011).
\newblock A survey of {DHT} security techniques.
\newblock {\em ACM Comp. Surveys}, 43(2).

\end{thebibliography}


\begin{thebibliography}{}

\bibitem[Boulic and Renault 1991]{boulic:91}
Boulic, R. and Renault, O. (1991).
\newblock 3d hierarchies for animation.
\newblock In Magnenat-Thalmann, N. and Thalmann, D., editors, {\em New Trends
  in Animation and Visualization}. John Wiley {\&} Sons ltd.

\bibitem[Knuth 1984]{knuth:84}
Knuth, D.~E. (1984).
\newblock {\em The {\TeX} Book}.
\newblock Addison-Wesley, 15th edition.

\bibitem[Smith and Jones 1999]{smith:99}
Smith, A. and Jones, B. (1999).
\newblock On the complexity of computing.
\newblock In Smith-Jones, A.~B., editor, {\em Advances in Computer Science},
  pages 555--566. Publishing Press.

\end{thebibliography}
